\documentclass[review]{elsarticle} 
\usepackage[hyphens]{url}
\usepackage{lineno} 
\linenumbers 
\biboptions{sort&compress} 
\usepackage{graphicx}
\usepackage{booktabs} 
\makeatletter
\def\ps@pprintTitle{%
 \let\@oddhead\@empty
 \let\@evenhead\@empty
 \def\@oddfoot{\it \hfill\today}%
 \let\@evenfoot\@oddfoot}
\makeatother

\usepackage[T1]{fontenc}
\usepackage{lmodern}
\usepackage{amssymb,amsmath}
\usepackage{ifxetex,ifluatex}
\usepackage{fixltx2e} 
\IfFileExists{upquote.sty}{\usepackage{upquote}}{}
\ifnum 0\ifxetex 1\fi\ifluatex 1\fi=0 
  \usepackage[utf8]{inputenc}
\else 
  \usepackage{fontspec}
  \ifxetex
    \usepackage{xltxtra,xunicode}
  \fi
  \defaultfontfeatures{Mapping=tex-text,Scale=MatchLowercase}
  
\fi
\IfFileExists{microtype.sty}{\usepackage{microtype}}{}
\usepackage{graphicx}
\makeatletter
\def\maxwidth{\ifdim\Gin@nat@width>\linewidth\linewidth
\else\Gin@nat@width\fi}
\makeatother
\let\Oldincludegraphics\includegraphics
\renewcommand{\includegraphics}[1]{\Oldincludegraphics[width=\maxwidth]{#1}}
\ifxetex
  \usepackage[setpagesize=false, 
              unicode=false, 
              xetex]{hyperref}
\else
  \usepackage[unicode=true]{hyperref}
\fi
\hypersetup{breaklinks=true,
            bookmarks=true,
            pdfauthor={},
            pdftitle={No early warning signals for stochastic transitions: insights from large deviation theory},
            colorlinks=true,
            urlcolor=blue,
            linkcolor=magenta,
            pdfborder={0 0 0}}
\begin{document}
\begin{frontmatter}
  \title{No early warning signals for stochastic transitions: insights from large
         deviation theory}
  \author[cstar]{Carl Boettiger\corref{cor1}}
  \cortext[cor1]{Corresponding author}
  \ead{cboettig@ucsc.edu}
  \address[cstar]{Center for Stock Assessment Research, Department of Applied Math and Statistics, University of California, Mail Stop SOE-2, Santa Cruz, CA 95064, USA}
  \author[esp]{Alan Hastings}
  \address[esp]{Department of Environmental Science and Policy, University of California, Davis, CA, 95616 United States}
 \end{frontmatter}

In Boettiger \& Hastings {[}1{]} we demonstrated that conditioning on
observing a purely stochastic transition from one stable basin to
another could generate time-series trajectories that could be mistaken
for an early warning signal of a critical transition (such as might be
due to a fold bifurcation {[}2{]}), when instead the shift is merely due
to chance. While the goal was to highlight a potential danger in mining
historical records for patterns showing sudden shifts when seeking to
test early warning techniques, Drake {[}3{]} draws attention to a
potentially more interesting consequence of our analysis. Drake argues
that the bias observed could be used to forecast purely stochastic
transitions -- a task previously thought to be impossible {[}4{]}. We
feel this interpretation is too generous. The pattern Drake points to
arises in any large deviation, regardless of whether a system is or is
not at elevated risk for a transition. We illustrate this pattern in
systems with and without bistability, demonstrating that early warning
signals do not exist for purely stochastic transitions.

Here we provide a numerical demonstration that the pattern in question
for consideration of an early warning signal appears not only before
purely stochastic transitions (as seen in Reference 1) but during any
large deviation. As large deviations can occur even in stochastic
systems that have only a single stable point, these patterns cannot be
considered indicators of stochastic transitions. We demonstrate this in
two scenarios: first using the Allee model of alternative stable states
considered in Reference 1, Eqn 2.1 - 2.2 and Figure 2, and then in a
simple Ornstein-Uhlenbeck (OU) model which has only a single stable
state. Rather than condition on a stochastic transition having occurred
(as in Reference 1), we now condition on having merely observed a
sufficiently large deviation. It does not matter precisely what
``large'' deviation is considered, only that the larger the deviation
the more replicates or longer simulation times will be needed to
sufficiently populate the sample. We pick values such that we get a
sample of a few hundred large deviation events in a sample of 20,000
replicates.

The OU model is defined by a stochastic differential equation in which
there is only a single optimum whose strength is proportional to the
displacement,

\[ dX_t = - \alpha X_t dt + \sigma dB_t, \]

where the state $X_t$ oscillates around a stable point (at zero in these
arbitrary units), driven by Brownian noise $dB_t$ of intensity $\sigma$
and restorative force $\alpha$.

The analysis for each model proceeds exactly as in Reference 1: For each
model we generate 20,000 replicate time series. We condition upon only
those experiencing a deviation of size $L$ ($X \leq 250$ in the Allee
model and $X \leq -4$ in the OU model). For the sequence of observations
immediately leading up to the large deviation we compute the warning
signals of variance and autocorrelation over a sliding window of half
the length of the time-series, and we summarise the increasing or
decreasing trend observed in the variance and autocorrelation using
Kendall's $\tau$ rank correlation coefficient (all following the method
for early warning indicators outlined in Reference 5). We repeat this
analysis on the entire set of time-series under each model to obtain
null distributions for $\tau$ statistic.

\begin{figure}[htbp]
\centering
\includegraphics{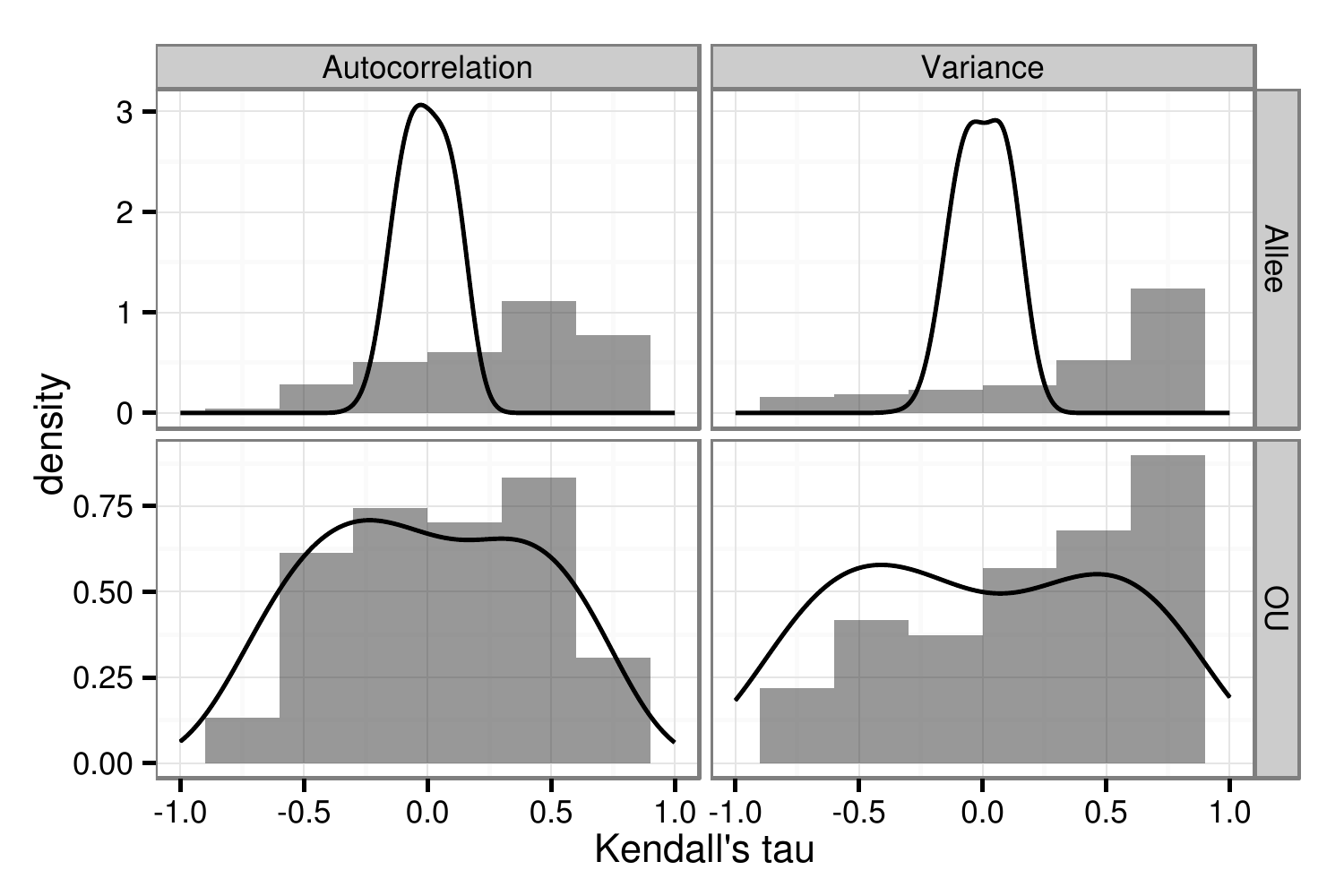}
\caption{Figure 1. Histogram shows the frequency the correlation
statistic $\tau$ observed for each warning signal (variance,
autocorrelation coefficient) on the large deviation samples from each
model. Background distribution of all samples show by smooth line
(kernel density estimate). More positive values of tau are supposed to
indicate a rising indicator which can be a signal of an approaching
transition {[}2{]}. The OU model uses $\alpha = 5$, $\sigma=3.5$,
$t \in (0, 10)$, 2000 replicates, 20,000 sample points each.
Conditionally selected trajectories experiencing a deviation of at least
-4, and analyzed the 1,500 data points prior to the threshold to
determine a warning signal (following Reference 5). (Code at:
https://raw.github.com/cboettig/earlywarning/resubmission/inst/doc/Figure1.Rmd,
data at:
https://raw.github.com/cboettig/earlywarning/resubmission/inst/doc/Figure1.csv
(Once published, code and data will appear in Dryad via doi:
10.5061/dryad.1dj62)}
\end{figure}

We find (Figure 1) that $\tau$ is significantly skewed towards positive
values when conditioning on large deviations in both models. This
demonstrates that it is the presence of the large deviation, not the
presence of the stochastic transition we condition on in Reference 1,
that is responsible for this pattern (just as we claimed without example
then).

Observing the bias shown in the figures here depends on having a rapid
enough sample frequency to capture the escape trajectory and a long
enough trajectory for the statistic to demonstrate an increase over
time. Since large deviations due to stochastic forces alone must be
fast, so must the accompanying warning signal and management response
(which will show up on the time scale of the perturbation). Note that
fast relative to the system dynamics may or may not be fast relative to
the timescale of management (just as with bifurcation-driven warning
signals, Reference 6). The wider null distribution in the OU model
results from the sample window being shorter relative to the system
timescale.

One might consider this a corollary of the Prosecutor's Fallacy we
originally presented, which demonstrated that examples of sudden
transitions historically selected from the literature could be mistaken
for positive evidence of early warning signals when they were in fact
due to purely stochastic transitions. Here we have seen how any large
deviation could be similarly misleading, whether or not it results in a
stochastic transition to an alternative stable state. From a classical
result of the large deviation theory one can gain considerable intuition
about why these chance deviations show much higher variance and
autocorrelation than expected from the stationary distribution of a
stable point. Though large deviations are rare -- the time we must wait
to observe a deviation of size $L$ in the system above scales as
$\exp\left(L^2/\sigma^2\right)$ (the familiar Arrhenius relationship),
when these deviations occur they occur very rapidly. The expected time
for an excursion to a distant point $L$ that does not again cross the
stable point before reaching $L$ scales as $\log(L/\sigma)$, just as a
trajectory returning down the gradient of the attractor from $L$ to the
stable point (proofs in Reference 7 or Reference 8). While most
trajectories in the stationary distribution take steps in each direction
with equal probability, these large deviations moving rapidly to the
boundary will consequently show the greater autocorrelation. In
achieving a much greater deviation than typically observed, these
trajectories will also show an increase in variance, as observed. That
such trajectories appear to be pulled in the direction of their escape
rather than climbing away against a restorative force has led to
confusion before. Reference 8 argues how this shows how a ``punctuated
equilibrium'' pattern of stasis followed by rapid change could arise
entirely from small steps, and Reference 9 empirically demonstrates this
phenomenon in the trajectories of local population extinctions.

In conclusion, we heartily agree with the need for a decision-theoretic
approach to early warning signal questions {[}10{]}. Central to a
decision-theoretic approach is enumerating alternative scenarios that
are possible given the observed data. We have highlighted how purely
stochastic transitions and large deviations are such possibilities. The
challenge of sufficient or unique early warning indicators is not
limited to stochastic shifts, but includes the more typical critical
transitions. For instance, rising variance or autocorrelation patterns
typical of fold bifurcations can be observed in more benign bifurcations
or smooth transitions {[}11{]}. Early warning signals may offer a
promising technique that will one day allow us to avoid seemingly
unpredictable catastrophes -- but we must not lose sight of just how
difficult are the challenges involved. A key step here and for early
warning indicators more generally is to understand these other
circumstances in which they can arise, that we may then develop ways to
eliminate those possibilities. Though we may never be able to detect
purely stochastic transitions, perhaps these approaches in this
discussion may lead to more unique and sufficient indicators for true
critical transitions.

The authors acknowledge the generous support of NSF grant EF 0742674 to
AH and helpful comments from TA Perkins, an anonymous reviewer and
reviewer P. Ditlevsen.

1 Boettiger, C. \& Hastings, A. 2012 Early warning signals and the
prosecutor's fallacy. \emph{Proceedings of the Royal Society B:
Biological Sciences} (doi:10.1098/rspb.2012.2085)

2 Scheffer, M. et al. 2009 Early-warning signals for critical
transitions. \emph{Nature} \textbf{461}, 53--9.

3 Drake, J. M. 2013 Early warning signals of stochastic switching.
\emph{Proceedings of The Royal Society B} \textbf{in press}.

4 Ditlevsen, P. D. \& Johnsen, S. J. 2010 Tipping points: Early warning
and wishful thinking. \emph{Geophysical Research Letters} \textbf{37},
2--5. (doi:10.1029/2010GL044486)

5 Dakos, V., Scheffer, M., van Nes, E. H., Brovkin, V., Petoukhov, V. \&
Held, H. 2008 Slowing down as an early warning signal for abrupt climate
change. \emph{Proceedings of the National Academy of Sciences}
\textbf{105}, 14308--12. (doi:10.1073/pnas.0802430105)

6 Hughes, T. P., Linares, C., Dakos, V., van de Leemput, I. a \& van
Nes, E. H. 2013 Living dangerously on borrowed time during slow,
unrecognized regime shifts. \emph{Trends in ecology \& evolution}
\textbf{28}, 149--55. (doi:10.1016/j.tree.2012.08.022)

7 Ludwig, D. 1981 Escape from Domains of Attraction for Sytstems
Perturbed by Noise. In \emph{Nonlinear Phenomena in Physics and Biology}
(eds R. H. Enns B. L. Jones R. M. Miura \& S. S. Rangnekar), pp.
549--566. Boston, MA: Springer New York.

8 Lande, R. 1985 Expected time for random genetic drift of a population
between stable phenotypic states. \emph{Proceedings of the National
Academy of Sciences} \textbf{82}, 7641--7645.

9 Drake, J. M. \& Griffen, B. D. 2009 Speed of expansion and extinction
in experimental populations. \emph{Ecology letters} \textbf{12}, 772--8.
(doi:10.1111/j.1461-0248.2009.01325.x)

10 Boettiger, C. \& Hastings, A. 2012 Quantifying limits to detection of
early warning for critical transitions. \emph{Journal of The Royal
Society Interface} \textbf{9}, 2527--2539. (doi:10.1098/rsif.2012.0125)

11 Kéfi, S., Dakos, V., Scheffer, M., Van Nes, E. H. \& Rietkerk, M.
2012 Early warning signals also precede non-catastrophic transitions.
\emph{Oikos} (doi:10.1111/j.1600-0706.2012.20838.x)

\end{document}